\documentclass{article}

\usepackage{microtype}
\usepackage{graphicx}
\usepackage{subfigure}
\usepackage{booktabs} %

\usepackage{hyperref}
\usepackage{multirow}
\usepackage{adjustbox}

\usepackage[accepted]{icml2024}

\usepackage{amsmath}
\usepackage{amssymb}
\usepackage{mathtools}
\usepackage{amsthm}
\usepackage{tabularx}
\usepackage{bbding}
\usepackage[subtle]{savetrees}

\usepackage[capitalize,noabbrev]{cleveref}

\theoremstyle{plain}

\theoremstyle{definition}

\theoremstyle{remark}

\usepackage[textsize=tiny]{todonotes}
\usepackage[font=small]{caption}


\DeclareMathSizes{10}{9}{6}{5}

\renewcommand{\paragraph}[1]{\vspace{0.2em}\noindent\textbf{#1$\quad$}}

\icmltitlerunning{An Efficient Self-Learning Framework For Interactive Spoken Dialog Systems}

\begin{document}

\twocolumn[
\icmltitle{An Efficient Self-Learning Framework For Interactive Spoken Dialog Systems}

\icmlsetsymbol{equal}{*}

\begin{icmlauthorlist}
\icmlauthor{Hitesh Tulsiani}{equal,amazon}
\icmlauthor{David M. Chan}{equal,amazon,berkeley}
\icmlauthor{Shalini Ghosh}{equal,amazon}
\icmlauthor{Garima Lalwani}{amazon}
\icmlauthor{Prabhat Pandey}{amazon}
\icmlauthor{Ankish Bansal}{amazon}
\icmlauthor{Sri Garimella}{amazon}
\icmlauthor{Ariya Rastrow}{amazon}
\icmlauthor{Björn Hoffmeister}{amazon}
\end{icmlauthorlist}

\icmlaffiliation{amazon}{Amazon AGI}
\icmlaffiliation{berkeley}{UC Berkeley (work done while at Amazon)}
\icmlcorrespondingauthor{Shalini Ghosh}{ghoshsha@amazon.com}

\icmlkeywords{}

\vskip 0.3in
]

\printAffiliationsAndNotice{\icmlEqualContribution} %

\begin{abstract}
Dialog systems, such as voice assistants, are expected to engage with users in complex, evolving conversations. Unfortunately, traditional automatic speech recognition (ASR) systems deployed in such applications are usually trained to recognize each turn independently and lack the ability to adapt to the conversational context or incorporate user feedback. In this work, we introduce a general framework for ASR in dialog systems that can go beyond learning from single-turn utterances and learn over time how to adapt to both explicit supervision and implicit user feedback present in multi-turn conversations. We accomplish that by leveraging advances in student-teacher learning and context-aware dialog processing, and designing contrastive self-supervision approaches with Ohm, a new online hard-negative mining approach. We show that leveraging our new framework compared to traditional training leads to relative WER reductions of close to 10\% in real-world dialog systems, and up to 26\% on public synthetic data.
\end{abstract}

\section{Introduction}
\label{introduction}

Automatic speech recognition (ASR) for dialog systems has traditionally been a focused field, where the primary goal is to produce a text transcript for an utterance given the acoustic signal corresponding to that utterance \cite{radford2023robust, baevski2020wav2vec, hsu2021hubert, mitra:2023:unified}. While such systems have been largely successful, particularly in the domain of dialog systems and voice assistants (leading to word error rates below 2\% on the Librispeech benchmark \cite{radford2022robust}), in real-world applications such single-utterance systems have been shown to struggle with a long-tailed distribution of rare words, proper nouns, etc., leading to decreased user satisfaction with such systems \cite{schwarz2023personalized, kim2018dialog, chang2021context, chen2019joint, sathyendra2022contextual, wei2021attentive}.

\begin{figure}[t]
    \centering
    \includegraphics[width=\linewidth]{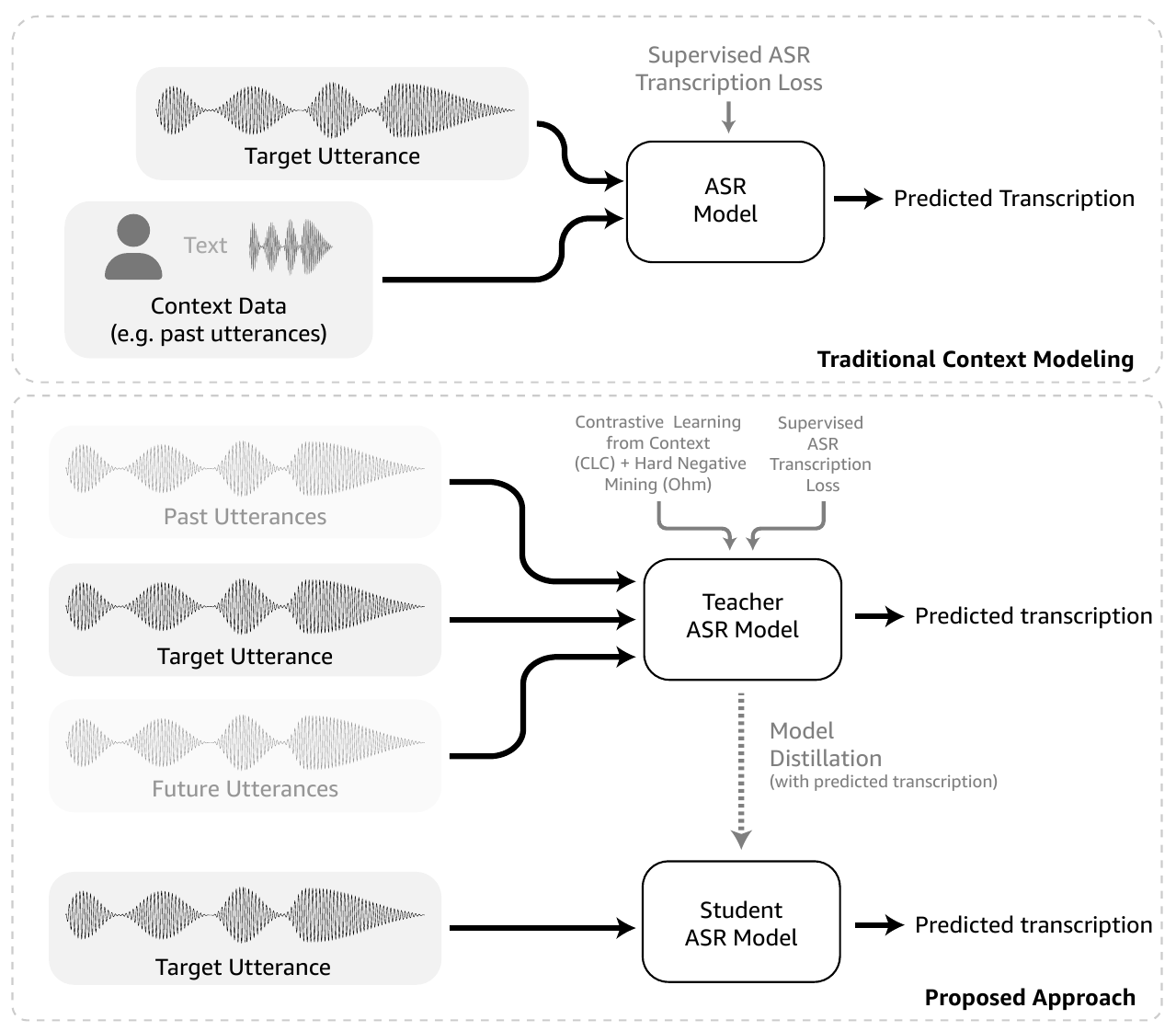}
    \caption{Traditional dialog systems learn to perform ASR using only supervised feedback with large-scale unsupervised/semi-supervised pre-training on single isolated utterances. This work introduces a novel general framework leveraging student-teacher distillation, contrastive learning, and online hard-negative mining, allowing ASR systems to learn from contextual clues and implicit feedback present in full conversational transcripts. Our two stage system naturally allows us to distill contextual signals from a context-aware teacher model to a context unaware student model.}
    \label{teaser}
\end{figure}

Such a struggle with long-tailed distributions has led to several promising directions of research aimed at specializing large-scale general models to handle rare words. These approaches generally center around fine-tuning where models are tuned on rare words as they are discovered \cite{li2022massively, Ku2024, gao2022wavprompt, chang2022speechprompt, yang2023english, hung2023low}, ``ASR model personalization'' where model parameters are locally adapted with user-specific context \cite{gourav2021personalization, biadsy2022scalable, shor2019personalizing}, or ``Contextual biasing'' where model inputs include additional user-specific context as part of the input to the model \cite{Jayanthi2023,kim2018dialog,tang2024improving, sathyendra2022contextual,chang2021context,chen2019joint,wei2021attentive,dingliwal2023personalization}. While these approaches have shown promising results, they often require additional compute during training or additional storage and retrieval for model parameter adapters, leading to significant compromises in terms of critical latency factors. These methods also often must rely on additional supervised training data during the training stage, leading to increased real-world system costs (such as data labeling) that are often not justifiable by marginal performance increases. 

In this work, we aim to address the two key challenges -- run-time performance and increased data costs -- by introducing a two-stage framework for context-aware automatic speech recognition. To reduce run-time performance costs, we explore the use of model distillation in a student-teacher framework — we leverage context signals during training of the teacher model, but do not use context signals during run-time inference of the student model for efficiency. To reduce data costs, we leverage recent advances in self-supervised learning from intrinsic contextual signals, augmented with a novel algorithm for online hard-negative mining — this enables the teacher model to learn context signals in a self-supervised fashion, eliminating the need for additional supervised data during training.

Evaluating the real-world performance of such a system is challenging, as there is little publicly available dialog data. To evaluate our approach, we run experiments on a large dataset of over 200K hours of real-world de-identified data from a popular conversational assistant system and show that leveraging context can help to significantly improve teacher model performance.

Our key contributions are as follows:
\begin{itemize}
    \item We introduce a multi-stage teacher model for automatic speech recognition in contextual dialog systems which is capable of leveraging both explicit context signals (through audio and text context) and implicit feedback signals (through contrastive learning combined with a novel online hard-negative mining algorithm) present in sequential task-oriented dialogues~\cite{chan2024task}.
    \item We leverage our teacher model in a distillation framework, and demonstrate that context signals can be distilled into a student model requiring no additional run-time compute compared to conventional systems.
    \item We demonstrate close to a 10\% relative WER improvement in real-world dialog systems applications for the teacher model, and up to 24.4\% WERR on the public OD3 dataset. Further, we demonstrate close to a 4\% relative WER improvement when our teacher model is distilled to the student model; providing strong evidence that learning from context at training time can be effective at test time (even when such context is unavailable). We additionally show our approach does extremely well in lower-resource domains, demonstrating up to a 22.8\% WERR on the tail distribution of real-world data.
\end{itemize}

\textbf{Terminology}  We refer to our system as "self-learning" to convey the system's ability to iteratively improve its performance by learning from dialogue contexts and user feedback. This goes beyond traditional SSL (self-supervised learning) techniques by integrating an "interactive" (though offline) component where the system learns from its environment rather than solely relying on pre-existing unlabeled data.

\section{Background \& Related Work}

Methods for modeling context for automatic speech recognition (ASR) systems can be generally categorized into two main categories: supervised methods, which rely on additional data and labels to infer context which is useful for speech recognition, and unsupervised methods, which learn context cues directly from the utterance, and any associated prior/future utterances. In this section, we discuss our proposed approach in context with prior approaches for context-aware ASR.

\subsection{Supervised Context Modeling}

As discussed in \autoref{introduction}, supervised context modeling for automatic speech recognition largely falls into three categories: 
\begin{itemize}
    \item \textbf{Fine-tuning}: where models are fine-tuned on specific datasets to increase global context awareness.
    \item \textbf{Model Personalization}: where model \textit{parameters} are updated on a per-user basis using a small set of user-specific samples.
    \item \textbf{Contextual biasing}: where models take additional context as input during the training and inference stages.
\end{itemize}

Each of these approaches has benefits and drawbacks. Perhaps the most common approach for context modeling is fine-tuning, which includes context by training on specialized datasets \cite{li2022massively, Ku2024, gao2022wavprompt, chang2022speechprompt, yang2023english, hung2023low}. Such an approach can be quite effective, as it turns a long-tail distribution problem into an in-domain problem. However, it requires the collection of explicit data for the target problem, and the scope of the context that a model can learn is limited to the collected data. Further, this data collection process is often expensive -- thus fine-tuning is often employed largely as an augmentation to an existing pre-trained model to fix specific errors, rather than as a good method for improving context awareness in general.

While fine-tuning adjusts the model globally to incorporate context (such as rare words), recently some approaches have been explored that focus on adjusting the model parameters locally to account for context. \citet{gourav2021personalization} show that small personalized models can be effective at incorporating information from user contexts, and \citet{biadsy2022scalable} show that small model adapters consisting of only a few thousand parameters can be locally fine-tuned for each user to improve ASR recognition performance. \citet{shor2019personalizing} show that such models can be trained using as little as five minutes of personalized speech. These approaches represent good ways of fine-tuning models to focus on users' individual needs, however, training model adapters for each user can be expensive and comes with storage requirements, inference performance questions, and data privacy concerns (as speech needs to be processed in the cloud, often with batches of other user data).

Instead of adjusting the model parameters, contextual biasing moves the inclusion of context to the input domain. Several types of context are effective including user information (such as contact names) \cite{tang2024improving, sathyendra2022contextual}, prior utterances \cite{chang2021context}, visual clues \cite{hsu2021hubert, chan2022avbert}, text catalogs \cite{dingliwal2023personalization, chan2023domain}. Outside of ASR, contextual biasing has long been shown to be effective in NLP applications \cite{novotney2022cue, shenoy2021contextual, zhao2019shallow, liu2017dialog, jaech2018personalized, kim2018dialog,  lin2015hierarchical, williams2018contextual, munkhdalai2022fast, sun2023contextual}. While contextual biasing represents an important component of context modeling, it is often limited by the requirement to collect supervised data during the training phase (i.e. contexts need to be explicitly collected and stored), as well as the requirement to have contexts during the inference phase, which can lead to significantly degraded performance in context-free scenarios. Further, contextual biasing often suffers from increased model complexity during inference, leading to slower response times and decreased user satisfaction.

\subsection{Unsupervised Learning From Dialogue Contexts}

Instead of learning context explicitly, unsupervised learning of context clues is a largely under-explored area in automatic speech recognition. Recent work has started to explore how we can learn contextual information from audio context alone. \citet{Hori2021AdvancedLE} and \citet{Hori2020TransformerBasedLE} take in several utterances at once, and use this joint context to perform automatic speech recognition on the final target utterance (demonstrating up to 15\% improvements in WER). Unfortunately, these methods require previous utterances to be available at test time and suffer when no previous context is available. Using only the target utterance, \citet{chan:interspeech:2022} show that other unrelated utterances within a batch can be used to filter noise from automated speech recognition models, however, they do not show that such methods help beyond global and local noise removal.

Instead of using audio, \citet{Kim2019GatedEI} apply BERT to the partial ASR transcript generated so far and use those BERT embeddings to inform the generation of the next token (effectively fusing the language model with the speech model). Similarly, both \citet{chang2023context} and \citet{duarte2024promptformer} show that taking in related text context from past utterances can improve ASR performance. These approaches, while interesting, focus primarily on text embeddings of prior context, and do not show that such ASR performance can persist in a context-free scenario (as is often the case in on-device learning) or discuss the inclusion of future context (available at train, but not inference time).

The approach of unsupervised learning from dialog contexts is closely inspired by \citet{chan2023domain} who introduce a family of methods (CLC) for learning from both past and future dialog contexts, using contrastive learning between the latent representations of past/future dialogues and the latent representation of the target utterance. The motivation behind this work is that audio that shares similar dialog contexts should have more similar latent representations, and thus, is more likely to contain relevant acoustic information. While our current approach borrows the PF-CLC objective from \citet{chan2023domain} as an additional pre-training objective on top of our fully supervised and self-supervised fine-tuning process, we also found that alone, PF-CLC led to only minor improvements in overall performance due to the small per-gpu batch sizes used during training (in our case, each GPU has a maximum batch size of 16). Thus, to improve the performance of PF-CLC in our real-world training scenario, we introduce a novel scheme for online hard-negative mining, allowing for improved efficiency when applying the CLC losses during fine-tuning. Further, \citet{chan2023domain} does not study in detail how to embed contexts during training, the impact of training on both past and future contexts, or if this context training persists under distillation. 

\subsection{Model Distillation}

Model distillation \cite{buciluǎ2006model, hinton2015distilling} has long been an effective tool when used to improve the performance of models during inference time. Not only are student models often more efficient than teacher models, but surprisingly, such models are often more effective on downstream test data \cite{radosavovic2018data, zhang2019your, pham2022revisiting}. These trends have held in ASR as well, where \citet{huang2018knowledge, kim2019knowledge, mun2019sequence} all show that large teacher ASR models can be distilled to resource-efficient but performant student models. 

Beyond model compression, however, model distillation has more recently also been used effectively to bridge streaming models and non-streaming models, as both \citet{yu2020dual} and \citet{kurata2020knowledge} have shown that using distillation between model architectures can lead to overcoming fundamental architecture limitations at inference time. Recently, \citet{futami2022distilling} showed that ASR can be improved by distilling in language models such as BERT, however, they did so using a vector-based representation, unlike our proposed approach that leverages model distillation and self-supervision \cite{pham2022revisiting}. Closest to our framework, \citet{masumura2021hierarchical} use distillation to bridge models that have long audio input contexts (such as several input utterances) to single utterance models, however their approach is limited to using text models in context, and they do not explore using audio context or learning the contextual hints from dialog.

\section{Self-Learning for Dialogue ASR}

An overview of our approach is given in \autoref{teaser}, and consists of two key components: a context-aware teacher model, leveraging both explicit context signals and implicit user feedback in the dialogue, and a single-utterance student model distilled from the context-aware teacher.

\subsection{Teacher Model}
\label{sec:teacher_model}

\begin{figure}[t]
    \centering
    \includegraphics[width=1.0\linewidth]{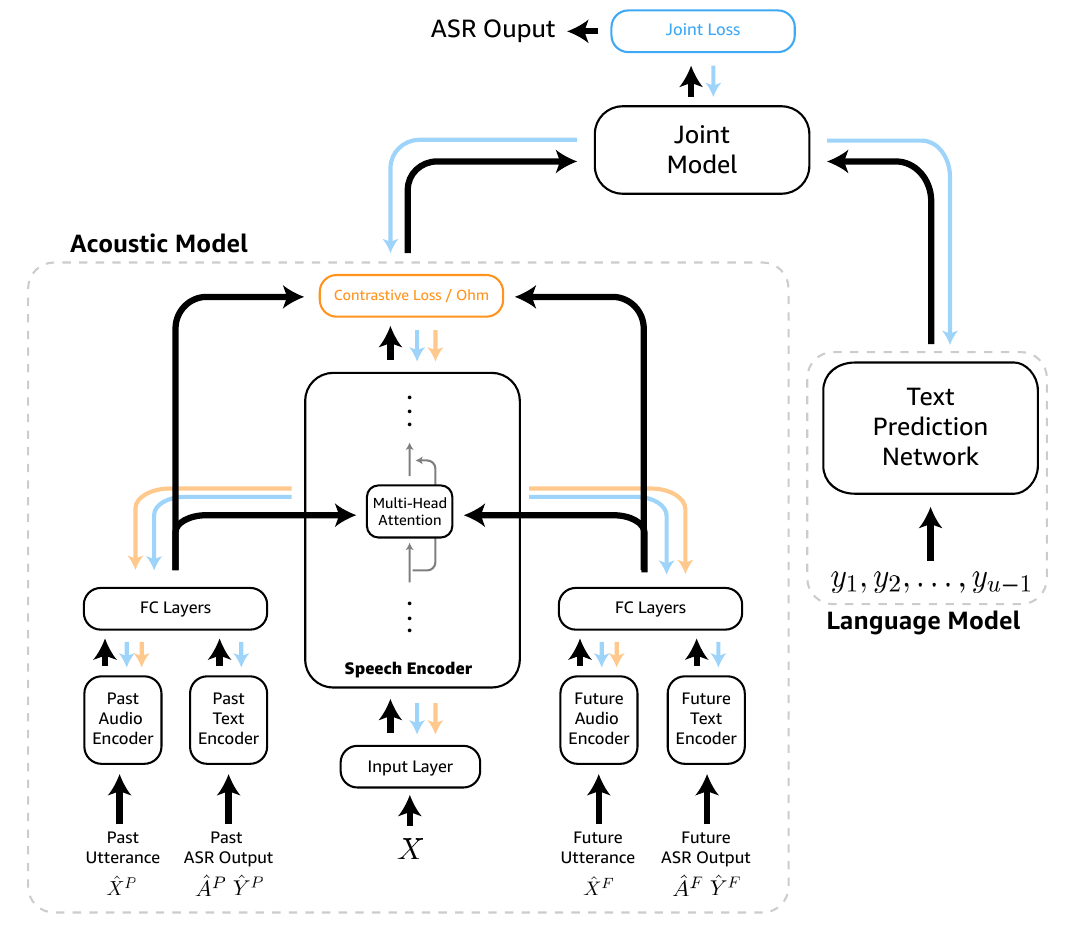}
    \caption{An overview of our approach. During training our teacher model ingests context from past/future audio and text along with the current utterance, and learns both implicitly and explicitly using CLC \cite{chan2024task} for implicit context learning and supervised joint loss for explicit learning from supervised data. \textcolor{black}{$\uparrow$} show data flow in forward-pass, and \textcolor{orange}{$\downarrow$} \textcolor{blue}{$\downarrow$} show loss propagation from each of the components.}
    \label{fig:teacher_model}
\end{figure}

Following best practices, our teacher model is composed of a Conformer-based transducer network \cite{graves2012sequence} - a nonstreaming model which can attend to all frames in an utterance. In addition, our teacher model also leverages both past and future contexts as explained in the following sections.

\subsubsection{Learning from Explicit Context}

In this work, we leverage several explicit context sources drawn from the dialogues themselves. The first is the audio context, formed by the sequence of user input queries in a given dialogue (preceding and succeeding audio context is represented as  $X^{P}$ and $X^{F}$ respectively in \autoref{fig:teacher_model}). The second is text context, the ASR one-best hypothesis (indicated as $\hat{Y}^{P}$ and $\hat{Y}^{F}$) corresponding to the sequence of user input queries in a dialogue along with the response generated by assistant encoded in text form (indicated as $A^{P}$ and $A^{F}$).

Traditionally, a transducer-based system, at each time step, outputs a probability distribution over its vocabulary (word-pieces) conditioned on the acoustic observations $X=x_1, x_2, ..., x_T$ and previously observed word-piece tokens $y_{1}, y_{2}, ..., y_{u-1}$, which could be expressed as $P(y_{u} | X, y_{1}, y_{2}, ..., y_{u-1})$. To model the explicit long-term context in the teacher model, we extend the above equation by further conditioning on the set of context signals, $Z = \{\hat{Y}^{P}$, $\hat{Y}^{F}$,
${A}^{P}$, ${A}^{F}$, ${X}^{P}$, ${X}^{F}\}$, to get $P(y_{u} | X, y_{1}, y_{2}, ..., y_{u-1}, Z)$.

\paragraph{Audio context modeling} Like \citet{Hori2021AdvancedLE} we explore two methods for adding audio context from dialogues:

\textit{Feature Concatenation:} In feature concatenation, we concatenate features of past and future utterances along with the seed utterance and pass it through the audio encoder. Encoder outputs are then segmented to extract embeddings corresponding to seed utterance and are combined with prediction network output to compute transducer loss. The presence of a self-attention network in the audio encoder allows us to learn the dependency on context streams.

\textit{Audio Embeddings:} For audio embeddings, past and future context is encoded via a separate encoder (called “context encoder”). We use either a HuBERT pre-trained conformer encoder or the audio encoder of the transducer network as the context encoder. These audio embeddings are passed through a multi-headed self-attentive pooling layer \cite{chang2023context} and concatenated in time dimension with keys and values of self-attention module (MHSA) in the audio encoder (represented in \autoref{fig:teacher_model}. Thus the inputs to MHSA (query - q, key - k, value - v) can be represented as $q = X; k = [{X}^{P}, X, {X}^{F}]; v = [{X}^{P}, X, {X}^{F}]$. This allows us to attend to the contextual signal on a per-query basis. Note here that the output and input of the MHSA module still have the same number of time frames. This ensures that no other component in the model needs to be modified. Another distinct advantage of re-purposing MHSA in this manner as opposed to introducing a separate cross-attention layer (to attend to context) is that it allows us to easily extend conventional single utterance models to be context aware.     %

\paragraph{Text context modeling} In addition to audio context, following \citet{duarte2024promptformer}, \citet{kim2019knowledge}, and \citet{chang2023context}, we explore two variants for encoding text context from prior utterances in a dialog:

\textit{BERT Embeddings: } In the BERT embedding case, we leverage a pre-trained text embedding model \cite{devlin2018bert} with 5M parameters based on BERT to generate a summary vector for the past/future text representations.

\textit{Learned Embeddings:} In the learned embedding case, text context is tokenized using a sentence piece model \cite{kudo2018sentencepiece} and each token is represented as a one-hot vector over vocabulary size. This is then converted to an embedding and combined in the self-attention layer of the audio encoder (similar to the audio embeddings described above).

\subsubsection{Learning from Implicit Context}
\label{sec:learning_from_implicit_context}
In a multi-turn interaction with a voice assistant, the user may repeat or rephrase their query (to correct the system) following an unexpected response by the assistant. To empirically establish that such user reformulations (implicit interactions) are correlated with ASR,
we conducted a simple experiment. We prepared two datasets: (i) Natural sampling: data is uniformly sampled to form our test set; and (ii) Reformulation sampling: we sample utterances that cause the user to repeat or rephrase their query. We then evaluate both our existing teacher and student models on these datasets. In this experiment, we observed that both the teacher and student models have significantly higher word error rates (11\% and 15\% respectively) on the reformulation sampling dataset compared to uniform sampling. This observation, combined with the fact that approximately 15\% of analyzed interactions had user reformulations, shows that user-provided implicit feedback can correct ASR errors. Please note that such feedback is not directly solicited through the dialogue but inferred from user corrections and follow-up queries, hence we refer to it as "implicit".

Thus, while it is possible to learn to leverage context signals from the explicit context, it is also important to learn from implicit signals in the data. Recently, \citet{chan2024task} showed that implicit context present in the dialogues can be used to further augment the training process through contrastive learning. Drawing on their work, in this work, we leverage the past-future CLC objective from \citet{chan2024task} (which we refer to as PF-CLC), to incorporate implicit context in addition to the explicit context discussed in the previous section.  In this PF-CLC approach, the positive pairs contain past/future/current utterances from the same utterance, while negative pairs are formed by past/future pairs from other utterances in the batches, further encouraging the model to organize the latent space semantically in addition to phonetically during pre-training. 

\subsubsection{Online Hard Negative Mining (Ohm)}

When training our self-learning based system, we found a significant correlation between the local GPU batch size and the performance of the pre-training~\cite{jain-etal-2024-multi}. We hypothesize that this correlation is caused by the PF-CLC learning introduced in \cite{chan2024task}, as with smaller local batch sizes, there are fewer ``hard negatives'' in each batch, leading to reduced efficiency when training with CLC-based losses. Unfortunately, scaling the local GPU batch size can be practically difficult, without introducing complex optimizers and training procedures for model sharding. This presents a challenging issue: how can we train contrastive models under restricted local batch sizes? 

While several technical methods have been developed for contrastive learning with small batch sizes including BASIC \cite{pham2023combined}, which leverage tools from gradient checkpoint and model parallelism to improve the ``effective" batch size, such methods have significant compute bottlenecks, and still rely on some form of all-reduce to compute the global contrastive loss. These all-reduces are technically complex, and on many GPU clusters can lead to significant communication overhead when machines must communicate with non-local devices.

Instead, of such a complex all-reduce based approach, we target the root of the problem by aiming to build more effective local batches (i.e. batches that will induce high contrastive loss by leveraging hard negative mining, introduced by \citet{robinson2020contrastive}. Traditionally, such methods for hard-negative mining rely on a pre-labeling step, where batches are pre-constructed in an offline-scan, and then consumed during training. Unfortunately, such a pre-labeling approach does not scale well as the size of the training data increases.  To remedy this, we introduce Ohm, a simple online hard-negative mining procedure that can run in line with traditional streaming data pipelines. An overview of the Ohm approach is given in \autoref{alg:ohm}. 

Ohm consists of several key stages each augmenting a data pipeline. In the first stage, samples are collected into an initial buffer $B$ using a stateful map. When the size of the initial buffer exceeds the update window size, then a parametric clustering method $C_\phi$ is fit on the samples from $B$. Note that this process happens per-device, and only with the samples yielded to each device, leading to non-blocking, and non-communicative behavior. Future version of Ohm could , however, communicate the $C_\phi$ model, leading to all-reduce like behavior with reduced overhead. Once $C_\phi$ had been trained, $C_\phi$ is used in a streaming fashion to assign labels to each sample. Finally, reservoir sampling \cite{vitter1985random} is used to sample batches of samples from each cluster group as they become available. This leads to batches which are generally closer semantically, even in the presence of a poor clustering algorithm $C_\phi$. Since the representations that we are using change, we periodically update $C_\phi$ (every 10,000 steps) using the buffer $B$. Ohm thus solves a practical problem: training on GPUs with less VRAM precludes the use of larger (more effective) batch sizes, and leveraging Ohm claws back some of that performance loss.

\begin{algorithm}[t]
\caption{Ohm: Online Hard Negative Mining}
\label{alg:ohm}
\begin{algorithmic}
\REQUIRE stream
\ENSURE Reordered stream

\medskip
\STATE Initialize state with zero index and an empty feature array
\medskip
\STATE  $\triangleright$ Step 1: Update clusters with streaming scan (stateful map)
\STATE stream $\gets$ stream.scan(initial\_state, update\_clusters)
\medskip
\STATE $\triangleright$ Step 2: Generate cluster labels for each sample
\STATE stream $\gets$ stream.map(generate\_labels)
\medskip
\STATE $\triangleright$ Step 3: Group samples by cluster label
\STATE stream $\gets$ reservoir\_sample(dataset)
\medskip
\STATE \textbf{return} batch\_processed dataset

\medskip
\STATE \textbf{procedure} update\_clusters(buffer, sample)
\STATE Add sample to buffer, and trim to update\_window\_size
\IF{buffer.size() $\geq$ update\_window\_size}
    \STATE Partially fit clusters to buffer
\ENDIF
\STATE \textbf{return} buffer

\medskip
\STATE \textbf{procedure} generate\_labels(sample)
\STATE Use clusters to assign sample a cluster label
\STATE \textbf{return} \{...sample, cluster\_label\}

\medskip
\STATE \textbf{procedure} reservoir\_sample(stream)
\STATE reservoir $\gets$ empty list for each cluster
\WHILE{stream.size() $>$ 0}
    \STATE sample $\gets$ stream.take()
    \IF{any reservoir[cluster].size() $>$ batch\_size}
        \STATE \textbf{yield from} reservoir[cluster]
    \ENDIF
\ENDWHILE

\end{algorithmic}
\end{algorithm}

\subsubsection{Reformulation Up-Sampling}

In addition to contrastive learning we further over-sample interactions containing reformulations during training of both the transducer and re-scorer. This approach can help to emphasize loss from reformulation samples, without introducing additional overhead. In our experiments, we empirically find an over-sampling ratio of 1:5 (1 reformulation to every 5 standard samples) to be effective (See \autoref{tab:reformulation}).

\begin{table}[t]
\scriptsize
\caption{\label{tab:reformulation} Word error rate reduction (WERR) on the \textsc{all} dataset when using different reformulation up-sampling rates.}
\begin{tabularx}{\linewidth}{Xccccccc}
\toprule
\textbf{Rate} & \textbf{None} & \textbf{1:10} & \textbf{1:5} & \textbf{1:4} & \textbf{1:3} & \textbf{1:2}  & \textbf{1:1} \\
\midrule
WERR & - & 0.23 & 0.46 & 0.42 & 0.39 & 0.35 & 0.24 \\
\bottomrule
\end{tabularx}
\end{table}

\subsection{Student-Teacher Distillation}
Our overall system comprises both a student and a teacher ASR system. Our student model is a Conformer-based transducer network \cite{gulati2020conformer} - a conventional streaming model i.e. it operates on single-utterance and attends to only past frames in the utterance. During run-time, governed by latency constraints, we use the student model to recognize user queries. These interactions (including reformulations and repeats) are captured -- details on how to determine reformulations are described in \autoref{sec:data}. Such interactions are then decoded using a context-aware teacher (discussed in the previous section) to get a recognition hypothesis, which acts as a label for semi-supervised training of student model \cite{hari2019lessons}.

\section{Experimental Design}
\label{sec:design}

In this section, we discuss the details used when evaluating our approach on both real-world conversational data, and open semi-synthetic data. 

\subsection{Datasets}
\label{sec:data}

All our transducer models are first pre-trained on 200k+ hours of de-identified ASR data (\textsc{pretrain}) using transducer loss without incorporating any contextual information. We then fine-tune and evaluate our approach on one of the two sources of data below: a closed-source real-world dataset from a conversational assistant, and the recently introduced OD3 dataset \cite{chan2024task}. 

\subsubsection{Closed Source Data}
\label{sec:closed_source_data}
This dataset consists of de-identified dialogues constructed from real-world interactions with a voice assistant. These dialogues are each constructed around a seed utterance, which is human transcribed. 
Given the seed utterance, the method pulls in all related conversations occurring 90 seconds before and after it. This step is iteratively applied, amassing additional conversational exchanges as they appear. If this approach yields over five utterances, the interval for gathering conversations is cut down to 45 seconds, and the process is repeated. This shortening of the interval persists until the number of utterances falls below five or the interval narrows to a 15-second minimum. These restrictions are enforced to ensure that utterances in a dialog are a semantically coherent interaction around the same request.

\paragraph{Mining dialogues with reformulations} To train and evaluate our system, we additionally detect a subset of dialogues consisting of reformulations. To detect reformulations, we use a text similarity-based approach. To be precise, we use cosine similarity and edit distance between the ASR hypothesis of the seed and context utterances (generated during the user's interaction with the assistant). The dialog is said to contain a reformulation if any \{seed, context\} pair has a similarity greater than the threshold. 

We \textbf{train} our context encoder teacher models on 10M de-identified dialogues. Additionally, we upsample dialogues containing user reformulations, by 20\%, during training of the transducer i.e. one in every five dialogues has reformulation. For \textbf{evaluation}, we only select dialogues containing reformulations and ensure that all utterances in the dialog are human-transcribed. By focusing on dialogues where the user reformulated his query, we ensure that the selected dialog has significant ASR errors (as discussed in section \autoref{sec:learning_from_implicit_context}) and where contextual signals are expected to be meaningfully related to user queries. We create two datasets for evaluation (1) \textsc{all}: All transcribed utterances across all validation dialogues (60K utterances) and (2) \textsc{ref}: A subset of \textsc{all} containing only utterances that lead to user reformulations of the query (8.5K utterances).

To \textbf{distill} the knowledge of the teacher model to the student model, we use closed-source dialogues containing reformulations. Dialogues used for distillation are distinct from those used for training teacher models and we don't require seed utterances to be human transcribed. Such dialogues are fed to a context-aware teacher model to obtain transcription for seed utterances. This single utterance audio-text pair (approximately 25,000 hours) is then used for training the student model. We refer to this as \textbf{s}emi-\textbf{s}upervised single utterance \textbf{r}eformulation \textbf{d}ataset (\textsc{ssrd}). 

\subsubsection{OD3}

Following \citet{chan2024task}, we further evaluate our models on the open directed dialogue dataset (OD3). The OD3 dataset is a semi-synthetic dataset, where human-generated task-oriented dialogues from several popular data sets are augmented with LLM-generated conversational errors and computer-generated TTS audio. OD3 contains 620K turns of audio (approximately 1,172 hours).

\subsection{Model Details}

For our experiments, we use transducer architecture, with Conformer \cite{gulati2020conformer} as the audio encoder. We experiment with two different teacher architectures: (i) 200M parameters: 17 conformer blocks and attention size of 1024 (ii) 1B parameters: 18 conformer blocks and attention size of 1536. Each conformer block is composed of four modules - multi-head attention and convolutional modules are sandwiched between two feed-forward modules. The convolutional module has a kernel size of 30 frames. Before conformer blocks, we use a pre-processing block consisting of two convolution layers, which takes in features at a 30ms frame rate, and has a kernel size of 5 and stride of 3.

Our student model has 1B parameters and consists of 18 conformer blocks with an attention size of 1536. 
However, for student models we restrict the attention module to only attend to past frames - this ensures user-perceived latency is minimal. For the prediction network, we use a two-layer LSTM network with 1024 hidden dimensions and a vocabulary of 4,000 tokens.  

\subsection{Training details}
\label{sec:training_details}

Our teacher model is pre-trained using the \textsc{pretrain} dataset for 500K iterations, using a per-gpu batch size ranging from 32 to 1, depending on the length of the sequence (sequences are batched to maximally use GPU memory) across either 64 P100 GPUs (for 200M model) or 64 A100 GPUs (for 1B model). We pre-train using an Adam optimizer - we linearly increase the learning rate for 5000 steps and thereafter decrease it proportionally to the inverse square root of the step (as per schedule described in \cite{vaswani2017attention}), and use magnitude-based gradient clipping with a value of $10$. 

We then fine-tune our teacher models for 150k steps, using an Adam optimizer with gradient clipping, featuring a learning rate decay schedule that starts at $1e^{-8}$, holds at $1e^{-5}$, and decays to $1e^{-6}$, with the clipping norm set to 0.3, and a schedule policy that adjusts the learning rate at 20K, 80K, and 600K training steps. In addition, we apply dynamic L2 regularization to the Multi-head self-attention layers of the conformer using a PiecewiseConstantDecay scheduler that increases the regularization factor at training steps 15K and 30K (calculated as $2$ $\times$ number of conformer blocks $\times$ warmup steps), with the regularization values set to $1e^{-6}$, $5e^{-6}$, and $1e^{-5}$ at these intervals. 

For models trained with contrastive learning, we use a set of $32$ learned clusters for hard-negative mining, with a buffer size of $4096$ for the online reservoir sampling. We leverage the BIRCH \cite{zhang1997birch} clustering algorithm, over the embeddings of $X^{p}$. In the future, we intend to explore additional clustering algorithms and leverage better distance functions for the Ohm mining approach. For the hyper-parameters of the PF-CLC loss, we follow the parameters in \citet{chan2024task}.
Our student model is pre-trained using \textsc{pretrain} dataset for 400K iterations with 64 A100 GPUs and a per-gpu batch size ranging from 128 to 1. For model distillation, we use both \textsc{ssrd} and \textsc{pretrain} data, sampled at differing ratios, and standard ASR transducer loss.

\section{Results and Discussion}
As discussed in \autoref{sec:design}, we evaluate our system on both closed-source data and the OD3 dataset. In general, we use both standard word error rate (WER, $\downarrow$) and relative word error rate improvement (WERR, $\uparrow$) to evaluate our system.

\subsection{Teacher Performance}

\begin{table}
\scriptsize
\caption{\label{tab:overall} WER improvements on the \textsc{all} and \textsc{ref} datasets for the teacher model. LE: Learned Embeddings, AE: Audio Embeddings, FC: Feature Concatenation, WERR: Word Error Rate Reduction. }
\begin{tabularx}{\linewidth}{lllXcc}
\toprule
\multirow{2}{*}{\textbf{Model}} & \textbf{Audio} & \textbf{Text} & \textbf{CLC +} & \multicolumn{2}{c}{\textbf{WERR ($\uparrow$)}} \\
& \textbf{Context} & \textbf{Context} & \textbf{Ohm} & \textsc{all} & \textsc{ref} \\
\midrule
Baseline (200M) & - & - & - & - & - \\
\multirow{8}{*}{Teacher (200M)} & FC & - & - & 3.73 & 7.04 \\
                                & AE/HuBERT & - & - & 1.94 & 4.04 \\
                                & AE/Transducer & - & - & 2.77 & 4.97 \\
                                & AE/Transducer & LE & - & 3.32 & 6.70 \\
                                & - & BERT & - & 0.69 & 2.66 \\
                                & - & LE & - & 2.63 & 5.66 \\
                                & FC & LE & -& 4.70 & 8.08 \\
                                & FC & LE & \tiny\Checkmark & \textbf{6.91} & \textbf{9.58} \\
\midrule
\multirow{2}{*}{Teacher (1B)} & - & - & - & 0.55 & 2.89  \\
                              & FC & LE & - & \textbf{5.53} & \textbf{9.24} \\
\bottomrule
\end{tabularx}
\end{table}

\begin{table}
\scriptsize
\caption{\label{tab:od3}  Results on OD3 (overall and repeat/rephrase inducing) using the 200M model. WER ($\downarrow$): Word Error Rate, BERT-S ($\uparrow$): Bert Score. B: Basline, CX: Context, C: CLC Loss, O: Ohm.}
\begin{tabularx}{\linewidth}{Xcccc}
\toprule
\textbf{Model} & \multicolumn{2}{c}{\textbf{Overall}} & \multicolumn{2}{c}{\textbf{Repeat/Rephrase}} \\
& \textbf{WER (WERR)} & \textbf{BERT-S} & \textbf{WER (WERR)} & \textbf{BERT-S} \\
\midrule
B                &  11.90 (-) & 0.9711 & 19.09 (-) & 0.9402 \\
B/CX             &  11.13 (6.47\%) & 0.9762 & 16.17 (15.29\%) & 0.9690 \\
B/CX/C           &  8.99 (24.4\%)  & 0.9812 & 13.81 (27.65\%) & 0.9737 \\
B/CX/C/O &       \textbf{8.73 (26.2\%)}  & \textbf{0.9817} & \textbf{13.21 (30.8\%)} & \textbf{0.9771} \\ 
\bottomrule
\end{tabularx}
\end{table}
\begin{table}
\scriptsize
\caption{\label{tab:od3_benchmark}  Zero-shot results on OD3 for several open-source models - Whisper \cite{radford2023robust}, Conformer \cite{gulati2020conformer}, Wav2Vec 2 \cite{baevski2020wav2vec}, Streaming Conformer \cite{tsunoo2021streaming}, CLC \cite{chan2024task}. Models in this table are not directly comparable (trained on differing data, setups, hyperparameters, optimizers etc.), but serve as a benchmark for performance on OD3 under several varying setups. WER ($\downarrow$): Word Error Rate, BERT-S ($\uparrow$): Bert Score}
\begin{tabularx}{\linewidth}{lXcXc}
\toprule
\textbf{Model} & \multicolumn{2}{c}{\textbf{Overall}} & \multicolumn{2}{c}{\textbf{Repeat/Rephrase}} \\
& \textbf{WER} & \textbf{BERT-S} & \textbf{WER} & \textbf{BERT-S} \\
\midrule
Whisper S (200M) & 11.24 & 0.9775 & 14.17 & 0.9727 \\
Whisper L (1.3B) & 8.51 & 0.9852 & 12.37 & 0.9792   \\
Conformer (100M, Librispeech) & 19.26 & 0.9612 & 22.19 & 0.9571\\
Wav2Vec 2 (433M, Librispeech) & 19.41 & 0.9582 & 22.03  & 0.9544 \\
Streaming Conformer (45M) & 14.38 & 0.9701 & 16.70 & 0.9665 \\
CLC (200M) & 8.99 & 0.9812 & 13.81 & 0.9737 \\
Our model (200M) & 8.73 & 0.9817 & 13.21 & 0.9771 \\
\bottomrule
\end{tabularx}
\end{table}

Our overall results for the teacher model on the \textsc{all} and \textsc{ref} are given in \autoref{tab:overall}. We can see that our system, combining the feature-concatenation audio context (\autoref{sec:teacher_model}), learned text context (\autoref{sec:teacher_model}), and CLC/Ohm losses (\autoref{sec:learning_from_implicit_context}), outperforms the baseline model by up to 9.58\% on the \textsc{ref} and up to 6.91\% on the \textsc{all} dataset. These trends hold across model sizes, as our context-enabled model has similar improvement in both the 200M and 1B cases, implying such improvements are model-size independent. Note that for ASR models, 1B parameters is generally considered quite large, given the challenging latency and run-time requirements for ASR applications. 

\paragraph{Impact of audio context:} Both the approaches of incorporating audio context (feature concatenation and audio embeddings) improve over a baseline system that doesn't use contextual signals. Improvements due to feature concatenation are larger ($> 7\%$ on the reformulation test set), which is not unexpected; by concatenating features we allow the model to attend to all context frames as necessary, as opposed to attending to summary vectors ($N=8$, in our experiments) coming from multi-headed attentive pooling. Among the two approaches of extracting audio embeddings, we find using a transducer encoder as a context encoder is marginally better, likely as the embeddings are ``on-policy'' for the trained model, as opposed to coming from an external embedding model.

\paragraph{Impact of text context:} In general, we find that encoding text via learned embeddings as opposed to summary vector by BERT encoder is more beneficial. This is aligned with the observation made above for audio context embeddings - again, likely because the latent space is ``on-policy'' and trained specifically on in-domain data. We also see that textual context under-performs audio context, likely due to incorrect text content from the teacher model.

\paragraph{Combining Context Types:} We get the best performance when both audio and text contexts are combined (compared to adding two modalities individually). Interestingly, the 200M model with context is significantly better than the 1B model without contextual signals, highlighting the efficacy of our proposed approach in modeling implicit context signals. On OD3, we can see that adding both context types leads to a 6.47\% performance improvement, tracking similarly to the performance improvements seen in the \textsc{all} dataset.

\begin{table}
\scriptsize
\caption{\label{tab:past_future_ablation} Ablation experiments on the teacher model (200M). WERR: Word Error Rate Reduction.}
\begin{tabularx}{\linewidth}{ccccXrr}
\toprule
\multicolumn{4}{c}{\textbf{Context Type}} & & \multicolumn{2}{c}{\textbf{WERR ($\uparrow$)}} \\
\textbf{Past} & \textbf{Future} & \textbf{Audio} & \textbf{Text} & & \textsc{all} & \textsc{ref} \\
\midrule
- & - & - & - &  & - & -  \\
\tiny\Checkmark & - & \tiny\Checkmark & - &   & 1.52 & 3.35  \\

\tiny\Checkmark & - & - & \tiny\Checkmark &  & 1.24 & 1.5  \\
\tiny\Checkmark & - & \tiny\Checkmark & \tiny\Checkmark &  & 2.90 & 5.08  \\
- & \tiny\Checkmark & \tiny\Checkmark & \tiny\Checkmark &  & 3.60 & 7.39  \\
\tiny\Checkmark & \tiny\Checkmark & \tiny\Checkmark & \tiny\Checkmark &  & \textbf{4.70} & \textbf{8.08}  \\
\bottomrule
\end{tabularx}
\end{table}

\begin{table}
\scriptsize
\caption{\label{tab:ohm} WER improvements on the \textsc{all} and \textsc{ref} datasets for the teacher model with CLC/Ohm and \textbf{context-aware teacher model as baseline}. Note: Baseline is different from Table 1 to ensure comparable training setup. WERR: Word Error Rate Reduction.}
\begin{tabularx}{\linewidth}{lllXcc}
\toprule
\multirow{2}{*}{\textbf{Model}} & \textbf{Context} & \textbf{CLC} & \textbf{Ohm} & \multicolumn{2}{c}{\textbf{WERR ($\uparrow$)}} \\
& &  &&  \textsc{all} & \textsc{ref} \\
\midrule
\multirow{3}{*}{Teacher (200M)}  & \tiny\Checkmark & - & - &  - & - \\
& \tiny\Checkmark & \tiny\Checkmark & - & 6.09 & 8.68 \\
 & \tiny\Checkmark & \tiny\Checkmark & \tiny\Checkmark  & \textbf{6.88} & \textbf{10.60} \\
\bottomrule
\end{tabularx}
\end{table}

\paragraph{Causal vs. Non-Causal Context:} In \autoref{tab:past_future_ablation}, we ablate the types of context that we show to the model. We observe that injecting non-causal (``future'') context \textit{during training} provides relative WERR of 7.39\% as opposed to 5.08\% on the \textsc{ref} dataset (as well as improvements on the \textsc{all} dataset), indicating that future context is significantly more important when correcting user reformulations. This is likely due to the fact that user reformulation is a ``future signal'' i.e. it follows the utterance that caused the error. 

\paragraph{Implicit Context Learning:} We can see that learning from the implicit context in the model is important for understanding and correcting dialog errors. As shown in \autoref{tab:ohm}, Adding CLC and Ohm to the baseline model leads to significant improvement in the overall performance, particularly on the \textsc{ref} dataset (so much that it enables a 200M parameter model to outperform a 1B parameter model without such losses). On the OD3 dataset (\autoref{tab:od3}), the performance is even more distinct, with learning from implicit context leading to up to a 26.6\% relative improvement over a baseline non-context model. In addition, zero shot comparison with other open source benchmark models is shown in \autoref{tab:od3_benchmark}.

\subsection{Distilling knowledge to student model}

\begin{table}
\scriptsize
\caption{\label{tab:distillation} Table showing the impact of distillation from a teacher model trained with implicit/explicit context. WERR: Word Error Rate Reduction. DE: Distillation Efficiency. }
\begin{tabularx}{\linewidth}{lcccc}
\toprule
\multirow{2}{*}{\textbf{Model}} & \multirow{2}{*}{\textbf{\textsc{ssrd} weight}} & \textbf{\% Params} & \multicolumn{2}{c}{\textbf{WERR ($\uparrow$) / DE ($\uparrow$})} \\
& & \textbf{Adapted} & \textsc{all} & \textsc{ref} \\
\midrule
Student (1B)  & - & - & - & - \\
\midrule
\multirow{3}{*}{+Distillation} & 100 & 20 & 1.38 / 19.97\% & \textbf{3.06 / 31.94\%} \\
                                & 20 & 20 & \textbf{1.76 / 25.47\%} & 1.2 / 12.52\%\\
                                & 50 & 100 & 1.51 / 21.85\% & 2.95 / 30.79\%  \\
\bottomrule
\end{tabularx}
\end{table} 

\autoref{tab:distillation} shows the performance of our model when distilled to a context-free student model. We can see that in all cases, the student model distilled from a context-trained model achieves superior performance. We also evaluate the distillation efficiency (DE) of the models -- how much of the WER gains of the teacher model were retained during distillation. It is interesting to note that when leveraging the  \textsc{ssrd} dataset, only 20\% of the parameters in the model are necessary during the distillation process to achieve the same WERR, compared to when less reformulation data is used (see \autoref{sec:training_details}), indicating that the using the pre-trained teacher model with context not only is more accurate, but can be more efficient as well.

\subsection{Tail-Distribution Performance}
\label{sec:tails}

\begin{table}
\scriptsize
\caption{\label{tab:tail}  WERR when normalized by the domain (instead of by-utterance) on the \textsc{all} dataset. WERR ($\uparrow$): Word Error Rate Reduction. SERR ($\uparrow$): Sentence Error Rate Improvement.}
\begin{tabularx}{\linewidth}{lllXcc}
\toprule
\textbf{Model} & \textbf{Context} & \textbf{CLC} & \textbf{Ohm} & \textbf{WERR} & \textbf{SERR} \\
\midrule
\multirow{3}{*}{Teacher (200M)} & \tiny\Checkmark & - & - & - & -  \\     
                                & \tiny\Checkmark & \tiny\Checkmark & - & 3.75 & 2.40 \\   
                                & \tiny\Checkmark & \tiny\Checkmark & \tiny\Checkmark & \textbf{6.64} & \textbf{7.79}  \\   
\bottomrule
\end{tabularx}
\end{table}

While overall WER is an important measure, many times, a strong indicator of user satisfaction is performance on a wide range of queries on different topics (Such as home automation, calling/messaging and shopping). In \autoref{tab:tail}, we present WERR and SERR (Sentence Error Rate Improvement) when the WER is computed on each topic independently, and then averaged instead of being averaged over all utterances (independent of domain, i.e. \autoref{tab:tail} makes the assumption that all domains are equally likely). From this, we can see that while our non-context models perform well on the most common utterances, the contrastive models lead to significant improvements in less-common domains in our \textsc{all} dataset, including queries categorized into shopping (82.86\% WERR), calling/messaging tasks (73.7\% WERR), and music request tasks (36.8\% WERR), all of which often need contextual disambiguation. On the other hand, while still in the long tail of the dataset, our approach performs worse than the baseline on home automation tasks (-22.68\% WERR), one of the less diverse tasks that requires less contextual disambiguation. In such cases, our model may be relying more on the context, than the target utterance: leading to decreased performance. It remains interesting for future work to explore how we can dynamically trade off between context clues (for challenging utterances), and non-context learning (for utterances that don't require contextual disambiguation).

\section{Conclusion}

This work introduces a framework that improves ASR in dialog systems through a dual-model approach to contextual learning: a context-aware teacher model that improves learning through explicit and implicit context signals, and a distilled student model that maintains efficiency during inference without context reliance. We achieve significant WER reductions, up to 9.58\% on real-world datasets and 26.6\% on the OD3 dataset, with the student model maintaining up to 33\% of the reduction without context across the distillation process. The enhancements observed, particularly for rare words and diverse user queries, indicate a path toward more robust and satisfying conversational experiences, notably, the pronounced gains for tail queries suggests that our approach can significantly improve performance on less common tasks. Future directions for this work involve exploring the dynamic adjustment of the relative importance of context versus the target utterance based on their predicted utility, error correction~\cite{Yang_2023} and safety~\cite{mehrabi_2023}. This could potentially unlock even greater improvements in ASR performance, paving the way for more intelligent and adaptable conversational AI systems.

\section*{Impact Statement}

This paper presents work whose goal is to advance the field of Machine Learning. There are many potential societal consequences of our work. Automatic speech recognition technology enhances accessibility, education, healthcare, legal processes, customer service, workplace productivity, language preservation, global connectivity, media accessibility, and safety across various societal sectors. While such impact is largely positive, it is important to recognize the impact of self-learning systems for automatic speech recognition on greater discussions in privacy and security, which are well discussed in related work \cite{Chennupati2022, aloufi2021configurable}.

\bibliography{main}
\bibliographystyle{icml2024}

\newpage
\appendix
\renewcommand\thefigure{\thesection.\arabic{figure}}  
\renewcommand\thetable{\thesection.\arabic{table}}    
\setcounter{figure}{0}  
\setcounter{table}{0} 
\onecolumn

\section{Insertions/Deletions/Substitutions in CLC/Ohm}
\label{app:tail2}

We can further break down the performance in \autoref{tab:ohm} in terms of insertions, deletions and substitutions, which is given in \autoref{tab:ohm_h}. We can see that adding CLC loss significantly improves the rate of deletion compared to baseline models. Unfortunately, this comes at the cost of improvement in substitution and insertion. CLC, instead of doing the best job of disambiguating generated tokens, focuses on recall as opposed to precision. Ohm improves the disambiguation, as at the cost of deletions: more tokens are dropped, but the tokens that are preserved are more accurate. 

\begin{table}[h]
\small
\caption{\label{tab:ohm_h}  WERR when normalized by the domain (instead of by-utterance) on the \textsc{all} dataset. WERR ($\uparrow$): Word Error Rate Reduction. SERR ($\uparrow$): Sentence Error Rate Improvement. INSR: Relative Insertion rate. SUBR: Relative Substitution Rate. DELR: Relative Deletion Rate}
\begin{tabularx}{\linewidth}{lllXccccc}
\toprule
\textbf{Model} & \textbf{Context} & \textbf{CLC} & \textbf{Ohm} & \textbf{WERR} & \textbf{SERR} & \textbf{SUBR} & \textbf{INSR} & \textbf{DELR} \\
\midrule
\multirow{3}{*}{Teacher (200M)} & \tiny\Checkmark & - & - & - & - & - & - & -  \\     
                                & \tiny\Checkmark & \tiny\Checkmark & - & 3.75 & 2.40 & 1.56 & 0.82 & 13.65 \\   
                                & \tiny\Checkmark & \tiny\Checkmark & \tiny\Checkmark & \textbf{6.64} & \textbf{7.79} & \textbf{3.19} & \textbf{1.70} & \textbf{6.18}  \\   
\bottomrule
\end{tabularx}
\end{table}

\end{document}